\documentclass[prl,twocolumn,letterpaper]{revtex4}
\usepackage[latin1]{inputenc}
\usepackage{amsmath,amsbsy}
\usepackage{amsfonts,hyperref}
\usepackage{bbm}
\usepackage{verbatim}
\usepackage{amsthm}

\newcommand{\ket}[1]{|#1\rangle}
\newcommand{\bra}[1]{\langle #1|}

\newtheorem{theorem}{Theorem}

\begin{document}
\author{Joseph M.~Renes$^1$ and Jean-Christian Boileau$^{2,3}$}
\affiliation{$^1$Institut f\"ur Angewandte Physik, Technische Universit\"at Darmstadt, Hochschulstr. 4a, 64289 Darmstadt, Germany\\
$^2$Perimeter Institute for Theoretical Physics, 35 King Street North, Waterloo, ON, N2J 2W9 Canada\\
$^3$Institute for Quantum Computation, University of Waterloo, Waterloo, ON, N2L 3G1 Canada}

\title{Privacy Amplification, Private States, and the Uncertainty Principle}

\begin{abstract}
We show that three principle means of treating privacy amplification in quantum
key distribution, private state distillation, classical privacy amplification, and 
via the uncertainty principle, are equivalent and interchangeable. By adapting the security 
proof based on the uncertainty principle, we construct a new protocol for private
state distillation which we prove is identical to standard classical privacy
amplification. Underlying this approach is a new characterization of private states,
related to their standard formulation by the uncertainty principle, which 
gives a more physical understanding of  
security in quantum key distribution.
\end{abstract}

\maketitle 
Privacy amplification is the art of extracting a secret key from a 
string which is partially-known to an eavesdropper~\cite{BBR88,BBCM95}.
In quantum key distribution (QKD) it plays a vital role as the
protagonists, Alice and Bob, would like  
to transform their shared, but not secret, raw key
into a verifiably secret key even when the eavesdropper
Eve has tampered with the quantum signals.

Heuristically, privacy amplification works by applying a suitable randomly-chosen
function to the raw key which scrambles and shortens it
so that Eve's limited knowledge of the input tells her nothing 
about the output. The canonical example is using a random public string
and computing the XOR with the original string. Provided Eve's information
is not too large, Alice and Bob can be confident that the
output will be secret. 

Broadly speaking, QKD has historically taken three   
main approaches to privacy amplification. Each is characterized by 
its treatment of the states held by the various 
parties to the protocol. The first focuses on 
Eve's marginal state conditional on the key string,
which she obtains in the course of eavesdropping.
Applying a random function to the key string results in new 
marginal states for Eve which 
are essentially identical. We term this method 
{\em classical privacy amplification} as it is an adaptation of 
privacy amplification against classical adversaries. It can be traced through
the sequence of papers~\cite{KMR05,RK05,CRE04,KGR,DW05}.

The remaining two approaches focus either 
concretely on the states held by Alice and Bob, including
any auxiliary systems, or abstractly on the key itself. In the former,
privacy amplification is recast as a virtual form of {\em private state distillation} 
in which Alice and Bob transform their initial shared quantum 
state into a private state, a state which yields secret keys upon 
measurement~\cite{HHHO05}.
Maximally entangled states are a subset of private states, so this 
method includes the techniques of applying entanglement distillation to privacy
amplification developed in~\cite{DEJMPS96,LoChau99,ShorPreskill00} and the subsequent
work employing the technique of Shor and Preskill. Means for distilling more general
private states were found in~\cite{RS06}.  

The latter approach of focusing
abstractly on the key itself, irrespective of its realization by either honest party
and disregarding any auxiliary systems not held by Eve, was employed in the
first QKD security proof by Mayers~\cite{Mayers96}, subsequently improved by
Koashi and Preskill~\cite{KP03}, and
finally culminated in a security proof based on the uncertainty principle by Koashi~\cite{K06}.
Here privacy amplification is viewed as a means of creating a virtual 
Pauli $X$ eigenstate and then obtaining the key by measuring the conjugate $Z$ observable.

In this letter we draw these three threads together and show they are equivalent
when privacy amplification is based on linear functions.
We do so by adapting Koashi's proof to give a new method of private 
state distillation and then prove it is identical to classical privacy amplification. 
The distillation technique follows from a new characterization of private states 
which is complementary to their standard description in the sense of the uncertainty
principle. This unifies various
approaches to the security of QKD, allowing the various means of treating
privacy amplification to be interchanged. Moreover, it provides a
more physical picture of how security arises from quantum mechanics. 
 
The new private state distillation method significantly generalizes that 
presented in~\cite{RS06}, 
which directly applied entanglement distillation techniques. 
Correction of phase errors afflicting the key subsystems
becomes easier for private states as the shield can store phase error information. 
Thus, not all phase errors need be corrected, increasing
the secret key yield above the entanglement yield.  
However, the resulting rates still do not always match those of
classical privacy amplification as the shared state is not always 
a classical mixture of states subjected to various phase errors.

Our results are presented as follows. 
We first show how the uncertainty principle inspires dual descriptions of private
states. Then the method of classical privacy amplification is shortly recounted 
before proceeding to the new approach to private state distillation. 
The details of the derivation of the secret key rate are presented from which 
the equivalence of the methods follows. Finally, we conclude with a view to
open problems and related issues.

\emph{Secret Keys and Private States.}---A 
perfect secret key shared by Alice and Bob is a uniformly-distributed random variable 
about which Eve has zero information. 
Thus a \emph{perfect secret bit} is defined as 
$\kappa_{ABE}=\left(\frac{1}{2}\sum_{k=0}^1 P^{k}_{A}\otimes P^k_B\right)\otimes \rho_E$
for any $\rho_E$, where $P^k=\ket{k}\bra{k}$. 

Private states are those quantum states for which independent measurements by
Alice and Bob yield a secret key. For secret \emph{bits}, our focus in the remainder
of the paper, these measurements might as 
well be standard basis measurements on the qubit key registers $A$ and $B$. 
The overall state can  be purified by including additional systems, be they 
\emph{shield} systems $S$ under the control of Alice and/or Bob or Eve's systems $E$.
A \emph{private state} $\gamma_{ABSE}$ is then a pure state of the form 
\begin{equation}
\frac{1}{\sqrt{2}}\sum_k \ket{kk}_{AB}V^{k}_S\ket{\xi}_{SE}=U_{ABS}\ket{\Phi}_{AB}\ket{\xi}_{SE},
\end{equation}
where the unitaries $V^{k}$ as well 
as the state $\ket{\xi}$ are arbitrary. The state $\ket{\Phi}$ is the canonical maximally-entangled 
state and the unitary $U_{ABS}=\sum_{j,k}P^{j,k}_{AB}\otimes V^{j,k}_S$ is called a \emph{twisting operator}. 

The fact that private states lead to secret keys and secret keys come from private 
states immediately follows~(cf.~Theorem 2 of~\cite{HHHO05b}). 
Measurement of a private state $\gamma_{ABSE}$ 
immediately yields $\kappa_{ABE}$ with
$\rho_E=\xi_{E}$. Conversely, suppose $\gamma_{ABSE}$ is a pure state yielding 
$\kappa_{ABE}$ under the prescribed measurement. It follows that
$\ket{\gamma}_{ABSE}=\frac{1}{\sqrt{2}}\sum_k \ket{kk}_{AB}\ket{\varphi^k}_{SE}$ for
some arbitrary normalized states $\ket{\varphi^k}$ and furthermore, that 
$\varphi_E^k=\rho_E$ for all $k$. Calling $\ket{\xi}_{SE}$ the purification of $\rho_E$,
we must have $\ket{\varphi^k}_{SE}=V^{k}_S\ket{\xi}_{SE}$ for some unitaries 
$V^{k}_S$ since all purifications of the same state are related by
unitaries on the purifying system. 
We have implicitly proven 
\begin{theorem}
A pure state $\gamma_{ABSE}$ is a private state if and only if 
(a) $p_{j,k}={\rm Tr}[\gamma_{ABSE}\,P^{j,k}_{AB}]=\frac{1}{2}\,\delta_{j,k}$, and (b) $\gamma_E^j=\gamma_E^{k}$ for all $j,k$. 
\end{theorem}
\noindent This formulation is straightforward: Eve can obtain no information
about the key when all her marginal states are identical. 
The approach of classical privacy
amplification is to prove the shared output state has this property. 

A different characterization of private states follows from considering
a hypothetical measurement by Alice in the $x$-basis. 
This produces conditional states of the $BS$ subsystem:
$\sigma^{x}_{BS}=2\,{_A}\bra{\widetilde{x}}\gamma_{ABS}\ket{\widetilde{x}}_A$,
where $\ket{\widetilde{x}}$ is the $x$th $x$-basis state. Then one has
\begin{theorem}
\label{thm:privstate2}
A pure state $\gamma_{ABSE}$ is a private state if and only if 
(a) $p_{j,k}={\rm Tr}[\gamma_{ABSE}\,P^{j,k}_{AB}]=\frac{1}{2}\,\delta_{j,k}$, and 
(b$'$) $\sigma^j_{BS}\,\sigma^k_{BS}=0$ for all $j\neq k$.
\end{theorem}
\begin{proof}
Suppose $\gamma_{ABSE}$ is a private state, for which condition (a) is satisfied by inspection. 
The states $BS$ conditional states are    
$\sigma_{BS}^x=Z^x_BU_{BS}\big(P^{\widetilde{0}}_B
\otimes \xi_{S}\big)U_{BS}^\dagger Z^x_B$,
where the unitary 
$U_{BS}{=}\sum_k P^k_B\otimes V^{k}_S$ for $P^{\widetilde{x}}=\ket{\widetilde{x}}\bra{\widetilde{x}}$,
and $Z^x$ is the $x$th power of $Z$, in contrast to all other upper indices appearing herein. 
Since $[Z_B,U_{BS}]{=}0$,
$\sigma_{BS}^x=U_{BS}\big(P^{\widetilde x}_B
\otimes \xi_{S}\big)U_{BS}^\dagger$, and  (b$^\prime$) follows immediately.

Conversely, by condition (a) we have $\ket{\gamma}_{ABSE}=\frac{1}{\sqrt{2}}\sum_k \ket{kk}_{AB}\ket{\varphi^k}_{SE}$. 
From the Schmidt decomposition $\ket{\varphi^k}_{SE}=\sum_{\ell}\sqrt{\lambda^k_\ell}\ket{\mu^k_\ell}_S\ket{\nu^k_\ell}_E$
define $Y^{k}_S{=}\sqrt{\varphi^k_S}V^{k}_S$ for unitary $V^k$ so that 
$\ket{\varphi^k}_{SE}=\sum_\ell Y^{k}_S\ket{\ell\ell}_{SE}$. 
Here $V^{k}{=}L^{k} (R^{k})^T$ using the unitaries
$L^{k}\ket{\ell}{=}\ket{\mu^k_\ell}$ and 
$R^{k}\ket{\ell}{=}\ket{\nu^k_\ell}$. Now we can write 
$\sigma^x_{BS}=Z^x_B\sigma_{BS}Z^x_{BS}$ for 
$\sigma_{BS}=\frac{1}{2}\sum_{jk}\ket{j}_B\bra{k}\otimes Y^{j}_S(Y^{k}_S)^\dagger$.
Condition (b$'$) then implies $(Y^{0})^\dagger\, Y^{0}= (Y^{1})^\dagger\, Y^{1}$. 
Defining $\ket{\xi}_{SE}{=}\sqrt{(Y^{0}_S)^\dagger\, Y^{0}_S}\sum_\ell \ket{\ell}_S\ket{\ell}_E$ we obtain $V^{k}_S\ket{\xi}_{SE}{=}\ket{\varphi^k}_{SE}$
and thus the operator $U_{BS}$ 
produces the private state: $\ket{\gamma}_{ABSE}=U_{BS}\ket{\Phi}_{AB}\ket{\xi}_{SE}$.
\end{proof}
We can understand the relationship between these 
two characterizations as an instance of the uncertainty principle,
which in entropic form requires that the sum of entropies of $x$- and $z$-basis 
measurements must not be less than unity~\cite{MU88}. 
Theorem 1 implies 
that Eve's entropy of Alice's $z$-basis measurement (i.e.~the key) is itself unity. 
Complementarily, theorem 2 means Bob's entropy 
(actually Bob and shield) of Alice's $x$-basis 
measurement is zero, so Eve's 
entropy of $z$ must be not less than unity.

\emph{Classical Privacy Amplification.}---An ideal 
privacy amplification protocol would output a perfectly secret key
key from the input of only partially secret data. This is 
too optimistic for practical applications however, and 
in this section we recapitulate 
the formulation of protocols which distill an \emph{approximately} 
secret key. We say $\rho_{ABE}$ is $\epsilon$-private when
$||\rho_{ABE}-\kappa_{ABE}||_1\leq 2\epsilon$. 
This definition ensures the key can be safely composed 
with any other cryptographic task and moreover, we can
interpret the definition as saying that the actual key $\rho_{ABE}$ is really the ideal 
key $\kappa_{ABE}$ with probability at least $1-\epsilon$~\cite{RK05,benor}~\footnote{In 
usual QKD protocols, the state obtained after bit error correction 
is close in trace distance to  $\psi_{ABE}^{\otimes n}$. Thus bit 
error correction and privacy amplification can be treated separately 
and the security of the whole QKD protocol ensured by the triangle
inequality.}.

Here we assume that the input to privacy amplification
is $\psi_{ABE}^{\otimes n}$, where $\psi_{ABE}=
\frac{1}{{2}}\sum_k P^{k,k}_{AB}\otimes \varphi^k_{E}$ describes
a shared but not necessarily secret bit. 
In QKD this product state is the product of a \emph{collective attack}
in which Eve tampers with each signal individually. 
More general coherent attacks have been dealt with by 
randomly permuting the quantum signals after receipt and then showing that
privacy amplification can extract the same key from the resulting
state as from a product state~\cite{GL03,R05}. 

Now, for $K$ the classical random variable 
held by Alice and Bob, and $I$ the quantum mutual information, one can  show 
\begin{theorem}[\cite{RK05, DW05}]
\label{thm:dwrate}
There exists a privacy amplification scheme to extract $n[1-I(K{:}E)]$ secret bits
from $\psi_{ABE}^{\otimes n}$, for $n\rightarrow\infty$. Moreover, this
is the maximum possible rate. 
\end{theorem}

The scheme in~\cite{DW05} works by 
selecting a function at random and 
applying it to each of the $A$ and $B$ systems; the output size of the
function is $n[1-I(K{:}E)]$ bits. The crux of that proof is a result on 
measure concentration, the generic term indicating when a random
variable is exponentially likely to be very close to its mean value. 
The random variable in this case is Eve's state $\varphi_E^\mathbf{k}$,
where $\mathbf{k}\in\{0,1\}^n$. Initially Eve's conditional states
are not close to the mean, but averaging over \emph{some} of
the $\mathbf{k}$ produces a new random variable which is. 
This partial average comes from regarding the random function 
as picking a random reversible function on 
the length-$n$ strings and then discarding 
(averaging over) 
the last $nI(K{:}E)$ bits. 

The privacy amplification function need not be completely random;
as shown in~\cite{RK05} any 2-universal family of hash functions suffice.
This includes random linear hashing, which we will use for private state distillation.

\emph{Private State Distillation.}---As 
in classical privacy amplification, the goal of private state distillation 
is to distill a state close to a private state, again measured by the 
trace distance. Since the key measurement is itself a quantum 
operation, an output state
$\epsilon$-close to a private state results in a key at least
$\epsilon$-close to $\kappa_{ABE}$.

Koashi's method
is to distill an $X$ eigenstate in a {\em single} abstract key register; its
immediate application to private states is obscured by the need to 
respect the form of the twisting operator. But by using a linear hash function
for privacy amplification we can neatly avoid this problem. 
The essential point remains that 
the honest parties have full information about an observable conjugate
to the key.

Initially Alice, Bob, and Eve share $\psi_{ABE}^{\otimes n}$, 
which can be purified using the shield system $S$ to the  state 
\begin{equation}
\label{eq:inputpsd}
\ket{\Psi}_{ABSE}=
\ket{\psi}^{\otimes n}_{ABSE}=\frac{1}{{2^n}}\sum_{\mathbf{k},\mathbf{x}}\ket{\widetilde{\mathbf{x}}}_AZ^\mathbf{x}_B\ket{\mathbf{k}}_{B}\ket{\varphi^\mathbf{k}}_{SE}.
\end{equation}
Generally, Bob cannot perfectly predict the outcome $\mathbf{x}'$ of 
Alice's hypothetical $x$-basis measurement since
his information is limited by the Holevo quantity $\chi$ of the 
ensemble $\mathcal{E}=\{\frac{1}{2},\rho_{BS}^x\}$, where 
$\rho_{BS}^x=2\,{_A}\bra{\widetilde{x}}\psi_{ABS}\ket{\widetilde{x}}_A$~\cite{K73}.
But then the distillation strategy suggests itself: have 
Alice provide Bob the missing information. If she narrows the possible 
$\rho^\mathbf{x}_{BS}$ to a suitably-random set of size
$2^{n\chi(\mathcal{E})}$, then the HSW theorem indicates that with high probability 
Bob can determine $\mathbf{x}'$~\cite{HSW}.

Having sketched the method roughly, we now turn to the details. 
Alice's announcement consists of the bits 
$h_i=\mathbf{u}_i\cdot\mathbf{x}'$
for $n[1{-}\chi(\mathcal{E})]$ randomly chosen $\mathbf{u}_i$, i.e.~a 
random linear hash of $\mathbf{x}'$. This can
be thought of as the result of measuring the observables $X^{\mathbf{u}_i}$,
which define Pauli $X$ operators for a set of ``encoded'' qubits. The complementary
subsystem of encoded qubits is associated with the set of $Z^{\mathbf{v}_j}$, 
where \mbox{$\mathbf{u}_i\cdot\mathbf{v}_j=0$} for all  $i,j$. 
Thus we can
decompose the space of Alice's (Bob's) physical qubit systems into virtual systems 
$A_1, A_2$ ($B_1, B_2$) corresponding
to the observables $Z^{\mathbf{v}_j}$ and $X^{\mathbf{u}_i}$, respectively. 
The post-announcement state is  
$\ket{\Psi'}_{A_1BSE}={}_{A_2}\bra{\widetilde{\mathbf{h}}}Z^\mathbf{h}_{B_2}\ket{\Psi}_{ABSE}$,
\begin{equation}
\ket{\Psi'}_{A_1BSE}=\frac{1}{2^{n\chi}}\sum_{\mathbf{y},\ell}\ket{\widetilde{\mathbf{y}}}_{A_1}Z^\mathbf{y}_{B_1}
\ket{\ell}_{B_1} \ket{\bar{\varphi}^\ell}_{B_2SE},
\end{equation}
where $\ket{\bar{\varphi}^\ell}_{B_2SE}=(2^{n(1{-}\chi)})^{-\frac{1}{2}}\sum_\mathbf{m}\ket{\mathbf{m}}_{B_2}\ket{\varphi^{(\ell,\mathbf{m})}}_{SE}$,
since Bob can apply $Z^\mathbf{h}_{B_2}$ after learning $\mathbf{h}$ from Alice. 
He is left to distinguish the states
$\varrho^\mathbf{y}_{BS}=Z^\mathbf{y}_{B_1}\Psi'_{BS}Z^\mathbf{y}_{B_1}$. 
Note that system $B_2$ has now become part of the shield. 

A slight modification of the HSW theorem ensures that with 
high probability 
the pretty good measurement~\cite{HW94} can distinguish the $\varrho^\mathbf{y}_{BS}$  
with arbitrarily small probability of error. The theorem originally applies to
the distinguishability of random subsets of $\rho_{BS}^\mathbf{x}$ and 
here we have a random subspace. However, in the Appendix we show that
the standard proof can be easily adapted to this case, and in fact more
generally to the use of 2-universal hashing. Bob's measurement 
has elements 
\begin{equation}
\label{eq:pgm}
E^\mathbf{y}_{BS}=\sqrt{T^{-1}_{BS}}\,\Big(\Pi_{BS}\,\Pi^{\mathbf{y}}_{BS}\Pi_{BS}\Big) \sqrt{T^{-1}_{BS}},
\end{equation}
for $T_{BS}=\sum_\mathbf{y}\Pi_{BS}\Pi^{\mathbf{y}}_{BS}\Pi_{BS}$,
and $\Pi^\mathbf{y}_{BS}$ ($\Pi_{BS}$) the projection onto the typical subspace of 
$\varrho^\mathbf{y}_{BS}$ ($\langle\varrho^{\mathbf{y}}_{BS}\rangle$),
the subspace spanned by eigenvectors whose eigenvalues 
are near the likely value. Here $\langle \cdot \rangle$ denotes the average value.

We can determine the 
$E^\mathbf{y}_{BS}$ explicitly and thereby obtain the twisting operator.  
Note that 
$Z^\mathbf{y}_{B_1}\varrho^{\mathbf{y}}_{BS}Z^\mathbf{y}_{B_1}=
\frac{1}{2^{n\chi}}\sum_{\mathbf{\ell},\ell'}
\ket{\mathbf{\ell}}_{B_1}\!\bra{\ell'}\otimes{\rm Tr}_{E}\big[\ket{\bar{\varphi}^\mathbf{\ell}}_{B_2SE}
\bra{\bar{\varphi}^{\ell'}}\big]$, 
meaning that system $B_2S$ determines typicality in both $\Pi_{BS}$ and $\Pi^\mathbf{y}_{BS}$. 
Following the proof of Theorem 2
we may then define $\bar{Y}^\mathbf{\ell}_{B_2S}$ so
that $\Pi_{BS}\Pi^{\mathbf{y}}_{BS}\Pi_{BS}$ becomes 
\begin{equation}
Z^{\mathbf{y}}_{B_1}\bigg(\sum_{\mathbf{\ell},\mathbf{\ell'}}
\ket{\mathbf{\ell}}_{B_1}\!\bra{\mathbf{\ell'}}\otimes \bar{Y}_{B_2S}^\mathbf{\ell}\bar{Y}_{B_2S}^{\mathbf{\ell}' \dagger}\bigg)Z^{\mathbf{y}}_{B_1}.
\end{equation}
Direct calculation gives $T_{BS}=2^{n\chi}\sum_\mathbf{\ell} P^\mathbf{\ell}_{B_1}
\otimes \bar{Y}_{B_2S}^\mathbf{\ell}\bar{Y}_{B_2S}^{\mathbf{\ell} \dagger}$ 
and the square root of the (pseudo) inverse follows.

Now consider the unitary $\bar{V}^\mathbf{\ell}$
which comes from the polar decomposition 
$\bar{Y}^\mathbf{\ell}=\sqrt{\bar{Y}^\mathbf{\ell}(\bar{Y}^\mathbf{\ell})^\dagger}\,\bar{V}^\mathbf{\ell}$; with it we can write
\begin{equation}
E^{\mathbf{y}}_{BS}=Z^{\mathbf{y}}_{B_1}\bigg(\frac{1}{2^{n\chi}}\sum_{\mathbf{\ell},\mathbf{\ell}'}
\ket{\mathbf{\ell}}_{B_1}\!\bra{\mathbf{\ell}'}\otimes \bar{V}_{B_2S}^\mathbf{\ell}\bar{V}_{B_2S}^{\mathbf{\ell}' \dagger}\bigg)Z^{\mathbf{y}}_{B_1}.
\end{equation}
Defining the $\bar{U}_{B S}=\sum_\mathbf{\ell}P^{\mathbf{\ell}}_{B_1}\otimes \bar{V}^{\ell}_{B_2S}$
we can express this in the more appealing form
$E^\mathbf{y}_{BS}=\bar{U}_{BS}
(P^{\widetilde{\mathbf{y}}}_{B_1}\otimes \mathbbm{1}_{B_2 S})
\bar{U}_{B S}^\dagger$. 

Thus, Bob's strategy is to 
untwist the shield as best he can and then measure his key system in the $x$-basis.
He and Alice obtain the same outcome with probability 
\begin{equation}
{\rm P}_s=\frac{1}{2^{2n\chi}}\sum_{\mathbf{\ell},\mathbf{\ell}'}
{}_{B_2SE}\bra{\bar{\varphi}^{\mathbf{\ell}'}}\bar{V}^{\mathbf{\ell}'}_{B_2S}\bar{V}_{B_2S}^{\mathbf{\ell} \dagger}
\ket{\varphi^{\mathbf{\ell}}}_{B_2SE}.
\end{equation}
If Bob can determine $\mathbf{y}$ with high probability, ${\rm P}_s\approx 1$, 
and $\bar{U}_{BS}^\dagger$ 
functions as an untwisting operator. Defining 
$\ket{\Psi''}_{A_1BSE}=\bar{U}_{BS}^\dagger\ket{\Psi'}_{A_1BSE}$, the 
squared fidelity of $\Psi''_{A_1B_1}$ with $\Phi_{A_1B_1}^{\otimes n\chi}$
equals ${\rm P}_s$. Then ${\rm P}_s\geq 1{-}\epsilon^2$ implies 
$||  \Psi''_{A_1B_1} - \Phi_{A_1B_1}^{\otimes n\chi}||_1\leq 2\epsilon$~\cite{FvG99} and therefore $\ket{\Psi'}_{A_1BSE}$ is $\epsilon$-private. 
 Altogether we have sketched a proof of 
\begin{theorem}
\label{thm:psd}
There exists a distillation procedure to distill $n\chi(\mathcal{E})$ private states 
from $\psi_{ABS}^{\otimes n}$ for $n\rightarrow\infty$.
\end{theorem} 
\noindent Note that this is the same rate found by Koashi.
Now the associated method of classical privacy amplification is simple. 
The key is the result of measuring $Z^{\mathbf{v}_j}$ 
which commutes with the private state distillation procedure. This key can just
as well be reconstructed from individual $Z$ measurements directly
and inherits privacy from the virtual procedure.

Theorems~\ref{thm:dwrate} and~\ref{thm:psd} 
give the secret key rates $1{-}I(K{:}E)$ and $\chi(\mathcal{E})$, corresponding
to distillation procedures following from the two descriptions of private states,
respectively.  Since these descriptions are equivalent, we expect the
associated distillation methods to have the same rate.  
This intuition can be confirmed either by direct calculation or by 
appealing to upper bounds applicable to either scenario. 
By the results of~\cite{RK05}, $\chi(\mathcal{E})\leq 1{-}I(K{:}E)$.
Conversely, $1{-}I(K{:}E)\leq\chi(\mathcal{E})$ or else by  
performing the classical privacy amplification coherently, 
as detailed in~\cite{DW05}, Bob would effectively be able to distinguish 
more of the states $\rho^\mathbf{x}_{BS}$ than possible.

\emph{Conclusions.}---We have found that the three 
principle means of treating privacy amplification are essentially
identical and interchangeable. The dual descriptions of private states on which
the respective distillation methods rest are shown 
to be elegantly related by the uncertainty 
principle. This provides an immediate and intuitive understanding of how the
quantum information about the key is balanced between the 
eavesdropper and shield and how the secret information can be extracted.

Care must be taken to incorporate these results into QKD security proofs. Here
Alice and Bob begin with a known state $\ket{\psi}_{ABSE}$, whereas
one of the main tasks of a key distribution protocol is to reliably estimate
the state shared by the various parties. The presence of a shield system makes 
this task more difficult, but recent work demonstrates how to estimate 
the parameters relevant to private state distillation~\cite{HHHLO06}.

Reduction of coherent attacks to the case of collective attacks studied here is
similarly intricate. This reduction has been
accomplished by creating a permutation invariant state $\vartheta^{(n)}_{ABE}$ 
by randomly scrambling the order 
of the quantum signals and then demonstrating that the chosen 
key distillation method produces just as many secret bits as 
from the product input $\psi^{\otimes n}_{ABE}$. 
It remains to be shown that when including the 
shield system this sort of reduction method still applies. In particular,
Bob must still be able to distinguish the $\rho^\mathbf{x}_{BS}$ even
though the $\mathbf{x}$ are no longer independently and identically distributed.
We will report on this in a future publication.

Finally, our result on achieving the Holevo bound using 2-universal hashing 
may be of independent  interest.

{\em Acknowledgements.}---We thank Gernot Alber, Hoi-Kwong Lo, 
Norbert L\"utkenhaus, and
Graeme Smith for helpful discussions. 
JMR recieved support from the Alexander von Humboldt foundation 
and the European IST project SECOQC, and JCB from NSERC and the Marie Curie foundation.

{\em Appendix}---Given a source described by the ensemble
$\mathcal{E}=\{p_x,\rho_x\}_{x=1}^d$ 
which distributes letters $x$ to Alice 
and states $\rho_x$ to Bob, we seek a 
protocol which enables Bob to learn $x$ and consumes 
few resources as possible. The idea is for Alice to send Bob
some (minimal) amount of information about $x$ so that he can 
then perform a measurement to distinguish between the quantum
states consistent with this information. 

2-universal hashing can be used for this purpose. 
A family of functions $f:\mathcal{X}\rightarrow \mathcal{Y}$ is 
2-universal if Pr$[f(x)=f(x')]\leq 1/|\mathcal{Y}|$ for all $x\neq x'\in\mathcal{X}$. 
Note that random linear hashing, as used in the main text, is 2-universal. 
Suppose Alice applies a random $f$ from the hash family to a block (length-$n$ string)
$\mathbf{x}$ of letters, using $\mathcal{X}=\{0,1,\dots,d{-}1\}^n$. 
Then Bob will be left to distinguish between the elements 
of $\{\rho_{\mathbf{y}}=\rho_{y_1}\otimes\dots\otimes\rho_{y_n}\,|\, f(\mathbf{y})=f(\mathbf{x})\}$, for 
which he uses the measurement $\{E_\mathbf{y}^f\}$ 
as defined in Eq.~\ref{eq:pgm}, with the slight change that $E_\mathbf{y}=0$ when $\mathbf{y}$ 
is nontypical. This rejects nontypical signals, which are in any case exceedingly rare.  
Adapting the presentation in Appendix B of~\cite{D05} shows this protocol
will have low error probability. 
We now specialize to $d=2$, but the argument is essentially the same
for the general case.

Given a function $f$, the average probability of error is given by
${\rm P}_{{\rm E}|f}=\sum_\mathbf{x}p_\mathbf{x}{\rm Tr}[\rho_\mathbf{x}
(\mathbbm{1}-E_\mathbf{x}^f)].$ Lemma 2 of~\cite{HN03} states
that $\mathbbm{1}-(S+T)^{-1/2}S(S+T)^{-1/2}\leq 2(\mathbbm{1}-S)+4T$
for $0\leq S\leq\mathbbm{1}$ and $T\geq 0$, which we can apply to $E_\mathbf{x}^f$
using $\Lambda_\mathbf{x}=\Pi\,\Pi_\mathbf{x}\,\Pi$ as $S$ and $\sum_{\mathbf{x}'\neq \mathbf{x}}\Lambda_{\mathbf{x}'}$ as $T$ to obtain
\begin{equation}
{\rm P}_{{\rm E}|f}\leq 2-2\sum_\mathbf{x} p_\mathbf{x} \Big({\rm Tr}[\rho_\mathbf{x}\Lambda_\mathbf{x}]-2\!\!\!\!\!\!\!\!\mathop{\sum_{\mathbf{x}'\neq \mathbf{x}}}_{f(\mathbf{x})=f(\mathbf{x}')}\!\!\!\!\!\!\!\!
{\rm Tr}[\rho_\mathbf{x}\Lambda_{\mathbf{x}'}]\Big),
\end{equation}
where $p_\mathbf{x}=p_{x_1}\!\!\cdots p_{x_n}$
When $\mathbf{x}$ is typical, 
Tr$[\rho_\mathbf{x}\Lambda_\mathbf{x}]\geq 1-3\epsilon$ 
and by construction $\Lambda_\mathbf{x}=0$
when $\mathbf{x}$ is not typical. Moreover,
the total probability of typical strings exceeds $1-\epsilon$, so
we obtain 
\begin{equation}
{\rm P}_{{\rm E}|f}\leq 8\epsilon+4\sum_{\mathbf{x}}p_\mathbf{x}
\!\!\!\!\!\!\!\!\mathop{\sum_{\mathbf{x}'\neq \mathbf{x}}}_{f(\mathbf{x})=f(\mathbf{x}')}\!\!\!\!\!\!\!
{\rm Tr}[\rho_\mathbf{x}\Lambda_{\mathbf{x}'}].
\end{equation}
Now average over the possible $f$: 
\begin{eqnarray}
{\rm P}_{{\rm E}}&\leq& 8\epsilon+4\sum_{\mathbf{x}}p_\mathbf{x}
\sum_{\mathbf{x}'\neq \mathbf{x}}{\rm Pr}[f(\mathbf{x}'){=}f(\mathbf{x})]\,
{\rm Tr}[\rho_\mathbf{x}\Lambda_{\mathbf{x}'}]\notag\\
&\leq&8\epsilon+\frac{4}{|\mathcal{Y}|}\sum_{\mathbf{x}}p_\mathbf{x}
\sum_{\mathbf{x}'\neq \mathbf{x}}{\rm Tr}[\rho_\mathbf{x}\Lambda_{\mathbf{x}'}]\notag\\
&\leq&8\epsilon+\frac{4}{|\mathcal{Y}|}\sum_{\mathbf{x},\mathbf{x}'}p_\mathbf{x}
{\rm Tr}[\rho_\mathbf{x}\Lambda_{\mathbf{x}'}]\notag\\
&=&8\epsilon+\frac{4}{|\mathcal{Y}|}\sum_{\mathbf{x}'}
{\rm Tr}[\rho^{\otimes n}\Lambda_{\mathbf{x}'}]
\end{eqnarray}
To evaluate the trace, note that ${\rm Tr}[\rho^{\otimes n}\Lambda_{\mathbf{x}'}]={\rm Tr}[\Pi\rho^{\otimes n}\Pi\Lambda_{\mathbf{x}'}]$. Since 
$\Pi\rho^{\otimes n}\Pi\leq 2^{-n[S(\rho)-\delta]}\Pi$ (Eq.~19 of~\cite{D05}) 
we have 
\begin{equation}
{\rm P}_{\rm E}\leq 8\epsilon+4\frac{2^{-n[S(\rho)-\delta]}}{|\mathcal{Y}|}
\sum_{\mathbf{x}'}{\rm Tr}[\Lambda_{\mathbf{x}'}].
\end{equation}
But again $\Lambda_{\mathbf{x}'}=0$ for nontypical $\mathbf{x}'$ while
 Tr$[\Lambda_{\mathbf{x}'}]\leq
2^{n[\sum_j p_j S(\rho_j)+\delta]}$ otherwise (Eq.~18), leading to
\begin{equation}
{\rm P}_{\rm E}\leq 8\epsilon+4\frac{2^{-n[\chi(\mathcal{E})-2\delta]}}{|\mathcal{Y}|}
\sum_{\mathbf{x}\in{\rm Typ}}1.
\end{equation}
Finally, the size of the typical set is less than $2^{n[H(p_i)+\delta]}$, so
putting it all together we have
\begin{equation}
{\rm P}_{\rm E}\leq 8\epsilon+4\,2^{n[H(p_i)-\chi(\mathcal{E})+3\delta]}
|\mathcal{Y}|^{-1}.
\end{equation}
By choosing $\log_2|\mathcal{Y}|=n[H(p_i)-\chi(\mathcal{E})+4\delta]$, the probability
of error can be made arbitrarily small. 

Since Bob ultimately learns $x$, an information
gain of $H(p_i)$ bits, but Alice only provides $H(p_i){-}\chi(\mathcal{E})$, the quantum
states themselves provide on average $\chi(\mathcal{E})$ bits, 
in accordance with the Holevo bound.


\begin{thebibliography}{99}
\bibitem{BBR88}C. H. Bennett, G. Brassard, and J.-M. Robert, 
SIAM J. Comput. {\bf 17}, 210 (1988).

\bibitem{BBCM95}  C. H. Bennett, G. Brassard, C. Cr\'epeau, and U. M. Maurer, 
IEEE Trans. Inf. Theory {\bf 41}, 1915 (1995). 

\bibitem{KMR05}R.~K\"onig, U.~Maurer, and R.~Renner, 
IEEE Trans. Inf. Theory {\bf 51}, 2391 (2005).

\bibitem{RK05}R. Renner and R. K\"onig, in \emph{Proceedings of the 
Second Theory of Cryptography Conference (TCC) 2005}, Lecture 
Notes in Computer Science, Vol. 3378 (Springer, Berlin, 2005), pp. 407-425.

\bibitem{CRE04}M. Christandl, R. Renner, and A. Ekert, quant-ph/0402131.

\bibitem{DW05} I.~Devetak and A.~Winter, Proc. Roy. Soc. Lond. A {\bf 461}, 
207 (2005). I.~Devetak and A.~Winter, Phys. Rev. Lett. {\bf 93}, 080501 (2004).  

\bibitem{KGR} R.~Renner, N.~Gisin, and B.~Kraus, Phys. Rev. A {\bf 72}, 012332 (2005).
B.~Kraus, N.~Gisin, and R.~Renner, Phys. Rev. Lett. {\bf 95}, 080501 (2005).

\bibitem{HHHO05}K.~Horodecki, M.~Horodecki, P.~Horodecki, and 
J.~Oppenheim, Phys. Rev. Lett. {\bf 94}, 160502 (2005).

\bibitem{DEJMPS96}D. Deutsch, A. Ekert, R. Jozsa, C. Macchiavello, 
S. Popescu, and A. Sanpera, Phys. Rev. Lett. {\bf 77}, 2818(1996). 

\bibitem{LoChau99} H.-K. Lo and H.~F. Chau, Science {\bf 283}, 2050 (1999).
 
\bibitem{ShorPreskill00} P.~W. Shor and J. Preskill, Phys. Rev. Lett. {\bf 85}, 441 (2000).

\bibitem{RS06}J.~M.~Renes and G.~Smith, Phys. Rev. Lett.~{\bf 98}, 020502 (2006).

\bibitem{Mayers96} D.~Mayers, Lect. Notes Comput. Sci. {\bf 1109} 343 (1996).

\bibitem{KP03}M.~Koashi and J.~Preskill, Phys. Rev. Lett. {\bf 90}, 057902 (2003).

\bibitem{K06} M.~Koashi, J.~Phys.~Conf.~Ser.~{\bf 36}, 98 (2006).

\bibitem{HHHO05b} K.~Horodecki, M.~Horodecki, P.~Horodecki, and 
J.~Oppenheim, quant-ph/0506189.

\bibitem{MU88} H.~Maassen and J.~B.~M.~Uffink, 
Phys. Rev. Lett.~{\bf 60}, 1103 (1988). 

\bibitem{benor} M. Ben-Or, M. Horodecki, D. W. Leung, D. Mayers, J. Oppenheim, 
Lect. Notes Comput. Sci. {\bf 3378}, 386 (2005).

\bibitem{GL03} D. Gottesman and H.-K. Lo, IEEE Trans. Inf. Theory {\bf 49}, 457 (2003).

\bibitem{R05} R.~Renner, Ph.D.~thesis, ETH, 2005; quant-ph/0512258.

\bibitem{K73} A. S. Kholevo, Probl. Peredachi Inf. {\bf 9} 177, (1973).

\bibitem{HSW} A.~S.~Holevo, IEEE Trans. Inf. Theory {\bf 44}, 269 (1998). 
B.~Schumacher and M.~Westmoreland, Phys. Rev. A {\bf 56}, 131 (1997).

\bibitem{HW94} P.~Hausladen and W.~K.~Wootters, J. Mod. Opt. {\bf 41}, 2385 (1994).

\bibitem{FvG99} C.~A.~Fuchs and J.~van~de~Graaf, 
IEEE Trans. Inf. Theory {\bf 45}, 1216 (1999).

\bibitem{HHHLO06} K.~Horodecki, M.~Horodecki, P.~Horodecki, D.~Leung, 
and J.~Oppenheim, quant-ph/0608195.

\bibitem{D05} I.~Devetak, IEEE Trans. Inf. Theory {\bf 51}, 44 (2005). 

\bibitem{HN03} M.~Hayashi and H.~Nagaoka, IEEE Trans. Inf. Theory {\bf 49}, 1753 (2003).

\end{thebibliography}
\end{document}